# A free-energy landscape picture and Landau theory for the dynamics of disordered materials

Mohammad Reza H. Javaheri

Department of Chemistry and Biochemistry, Arizona State University, Tempe, AZ 85287-1604

Ralph V. Chamberlin

Department of Physics, Arizona State University, Tempe, AZ 85287-1504

Abstract

Landau's theory of phase transitions is adapted to treat independently relaxing regions in complex systems using nanothermodynamics. The order parameter we use governs the thermal fluctuations, not a specific static structure. We find that the entropy term dominates the thermal behavior, as is reasonable for disordered systems. Consequently, the thermal equilibrium occurs at the internal-energy maximum, so that the potential-energy minima have negligible influence on the dynamics. Instead the dynamics involves normal thermal fluctuations about the free-energy minimum, with a time scale that is governed by the internal-energy maximum. The temperature dependence of the fluctuations yields VFT-like relaxation rates and approximate time-temperature superposition, consistent with the WLF procedure for analyzing the dynamics of complex fluids; while the size dependence of the fluctuations provides an explanation for the distribution of relaxation times and heterogeneity that are found in glass-forming liquids, thus providing a unified picture for several features in the dynamics of disordered materials.

In 1955, Williams, Landel, and Ferry introduced a procedure for characterizing the thermal and dynamic properties of polymer melts and other highly viscous glass-forming liquids.[1] The WLF procedure has found broad applications in areas such as the annealing and thermal tempering of glasses, the properties of lubricants in high-pressure contacts, and the stress, strain, and aging behavior of polymers.[2] The WLF procedure includes an equation for non-Arrhenius activation that is equivalent to the Vogel-Fulcher-Tammann[3-5] (VFT) law, which has long been used to describe the temperature dependence of the characteristic time scale in the relaxation of disordered materials. The procedure also utilizes time-temperature superposition for non-Debye relaxation, where the spectral width is assumed to be independent of temperature. Still there is no generally accepted explanation for these empirical features.[6] Over the past 15 years, key insight into the underlying mechanism has come from several experimental techniques, which have shown that the response of many complex systems involves nanometer-sized dynamically-correlated regions.[7-11] Especially relevant are the measurements and analysis that show *thermo*dynamic heterogeneity,[12-15] indicating that the regions are weakly coupled to the thermal bath and essentially uncorrelated with neighboring regions. Here we present a general approach for understanding WLF behavior that also yields this heterogeneity. Our approach is based on Landau's unified theory for phase transitions,[16,17] combined with Hill's theory of small-system thermodynamics[18-20] (nanothermodynamics) adapted to treat finite-sized thermal fluctuations inside bulk materials.[21-25] Our results indicate that a thermal transition, which is broadened by small-system fluctuations, underlies the dynamical liquid-glass transformation. The transition occurs when a mixture of competing states that is favored by entropy at high temperatures, shifts towards a larger fraction of the state that is favored by energy at low temperatures.

## BACKGROUND

Viscous liquids usually exhibit complex dynamics as a function of time (*t*) and temperature (*T*). Time-dependent relaxation is often characterized by the Kohlrausch-Williams-Watts (KWW) stretched exponential law $\Phi(t) \sim \exp[-(t/\tau)^\beta]$, with a stretching exponent $\beta$ that is usually less than one, in contrast to the simple Debye relaxation where $\beta=1$. Similarly, temperature-dependent activation is often characterized by the VFT super Arrhenius law $\tau \sim \exp[B/(T-T_0)]$, with relaxation times that usually diverge at a non-zero temperature $T_0$, in contrast to simple Arrhenius behavior where $T_0=0$. Now it is known from several experimental techniques that the net response comes from a distribution of dynamically correlated regions inside the sample. Typical regions have a length scale of 1-5 nanometers,[26-28] which corresponds to $N \sim 10^1$-$10^5$ "particles" (molecules or monomeric segments). The net response comes from the sum over all sizes,

$$\Phi(t) \propto \Sigma_N N g_N \exp[-t/\tau_N], \qquad \text{Eq. (1)}$$

where $g_N$ is the size distribution. Here the size-dependent relaxation time can be written in terms of the Arrhenius law $\tau_N \propto \exp[E_N/kT]$, where $E_N$ is the activation energy in the region. The challenge is to find a physical explanation for the behavior of $g_N$ and $E_N$ that is commonly found in the response from a wide range of complex materials.

Here we describe a unified approach to obtaining the distribution of activation energies that yield WLF behavior. Our main result is that the magnitude of the average internal energy per particle for *m* relaxing particles can be written as

$$|<\varepsilon_m>| = \frac{\varepsilon_2 kT}{m(kT - \varepsilon_2)}, \qquad \text{Eq. (2)}$$

where $\varepsilon_2$ is the second-order expansion coefficient. When used in the activation energy of the Arrhenius law $E_N \sim m\, |\langle\varepsilon_m\rangle|$, Eq. (2) provides an explanation for several features in the response of complex materials. The $kT$ term in the numerator of $|\langle\varepsilon_m\rangle|$ cancels the $kT$ term in the Arrhenius law to yield spectral widths that are nearly independent of temperature, giving approximate time-temperature superposition. The ($kT$-$\varepsilon_2$) term in the denominator of $|\langle\varepsilon_m\rangle|$ gives the VFT law. While the inverse size dependence yields a thermal equilibrium distribution of independently relaxing regions.

Our primary assumption is that the internal energy of a relaxing region can be characterized by a single parameter $L$, with $-1 \leq L \leq 1$. Thus, $L$ is a type of normalized order parameter (not a length), so that our assumption leads to an internal energy per particle of $\varepsilon(L)$. Because amorphous materials cannot be described by a single type of local structure, $L$ characterizes the dynamical state not a specific structure. For example, $L$ may parameterize a type of molecular vibration, such as those envisioned for the two-level systems that govern the thermal properties of glasses at low temperatures.[29] Alternatively, $L$ may parameterize molecular librations. Specifically, $L$ may correspond to the number of molecules in a region that have fluctuated clockwise about an axis, minus the number that have fluctuated counterclockwise, divided by the total number of molecules in the region. Thus $L=+1$ or $-1$ if all molecules librate in phase, while $L=0$ if their rotations are random. A more general interpretation is that $L$ parameterizes the relative fraction of two competing structures. For example, $L=+1$ may correspond to a bond-ordered structure having local symmetry, while $L=-1$ may correspond to a density-ordered structure having long-ranged symmetry, with $L=0$ for an equal mixture. Thus, the order parameter we use would be an interpolation between the two order parameters proposed by Tanaka[30]. Our theory is simplified by using a single order parameter for two competing structures, while still allowing mixtures that are entropically favored at high temperatures. In any case, our model is based on the assumption is that there is a single type of order parameter, with an average value <L> that is uniform across the sample, but with different values of L for each local region due to fluctuations about <L>.

The concept of an order parameter description of structural relaxation near the glass transition was introduced by Prigogine and Defay.[31] They showed that if the ratio

$$\Pi = \frac{\Delta C_p}{T_g V} \frac{\Delta \beta}{\Delta \alpha} \qquad \text{Eq. (3)}$$

is equal to one, then the configurational state of a homogeneous system is determined by a single internal parameter. Here $\Delta$ denotes the difference between the properties of the liquid and the glass, and $\alpha$, $\beta$, $C_p$, refer to the thermal expansion coefficient, isothermal compressibility, and isobaric specific heat, respectively. Davies and Jones[32] showed that if $\Pi>1$, the configurational state is described by more than one independent internal order parameter. Experimental data for a large number of polymeric and inorganic glasses show that $\Pi>1$ in almost every case,[33,34] which implies that the description of amorphous systems requires more than one order parameter.[35,36] However, Gupta and Haus[37] have concluded that even if $\Pi>1$, inhomogeneous systems can still be described by a single type of local order parameter, with a Prigogine-Defay ratio that deviates from unity proportional to the amount of heterogeneity. Thus, our approach of characterizing the thermal properties of viscous liquids in terms of a single parameter L is consistent with the experimental evidence for heterogeneity. An alternative interpretation comes from a recent frequency-independent analysis that might yield $\Pi=1$ for the glass transition.[38] If confirmed, this would imply that the local fluctuations do not significantly influence the average values of the thermodynamic properties in Eq. (3).

**THEORY**

The fundamental idea of Landau theory for phase transitions is to approximate the free energy of a system in a Taylor series expansion. For the sake of argument, we start at a more basic level by first neglecting entropy to expand just the internal energy per particle to second order, $\varepsilon(L)=-\frac{1}{2}\varepsilon_2 L^2$. Note that the zero point of energy is chosen so that the fully disordered system has $\varepsilon(0)=0$, and that the minus sign with $\varepsilon_2>0$ reflects the usual result that increasing the order lowers the energy of the system. At sufficiently low temperatures, higher-order terms in real systems could favor one of the ordered states, $L=+1$ or $L=-1$, but the $L^2$ term dominates the energy of most disordered systems. The solid curve in Fig. 1 shows that this energy as a function of $L$ forms a simple energy landscape, having two energy minima separated by a single energy barrier. The height of the barrier for a system of $N$ particles is $E_N=\frac{1}{2}N\varepsilon_2$. However, this constant activation energy gives the Arrhenius law, not the VFT law; and assuming a fixed distribution of sizes in Eq. (1) gives a spectral width that broadens with decreasing temperature, not time-temperature superposition. Similarly, unless some specific assumptions are made about the distribution of barrier heights, the usual picture of adding kinetic energy to a constant potential-energy landscape yields effective energy barriers that increase with decreasing temperature, causing further deviations from time-temperature superposition. Finally, there is no explanation for the independently relaxing regions that yield the heterogeneous dynamics. Thus, this simple expansion of the internal energy fails to describe any of the characteristic features in the dynamics of viscous fluids.

The next step towards a Landau theory is to add an entropy term to the internal energy to give the free energy. The free energy per particle expanded to fourth order is $f(L)=f_0+f_2L^2+f_4L^4$, where the parameters $f_0$, $f_2$, and $f_4$ may depend on $T$, but not on $L$. The probability of finding a particular configuration is proportional to $\exp[-mf(L)/kT]$. In the usual thermodynamic limit $m\to\infty$, the system is confined to the minimum free energy, yielding the usual Landau phase transition at the temperature where $f_2=0$. However, in this theory for long-ranged interactions with $m\to\infty$, there are no fluctuations and thus no dynamics, so that standard Landau theory is too simplistic to describe the dynamical properties of viscous liquids near the glass transition.

The main new feature in our theory is to consider finite-size effects. Specifically, we assume that a macroscopic sample is comprised of an ensemble of small regions, so that thermal fluctuations occur about the free-energy minimum of each small region. Although fluctuation effects can be neglected for the equilibrium behavior of macroscopic homogeneous systems, complex fluids have been shown to contain a heterogeneous distribution of independently relaxing regions, which via the fluctuation dissipation theorem implies a heterogeneous distribution of independently fluctuating regions. The ergodic theorem equates the thermal-equilibrium and time-averaged properties of these regions. From the symmetry of the fluctuations, the average value of the order parameter is zero above the transition, $<L>=0$, as in classical Landau theory. However, fluctuations give a non-zero value for the order parameter squared

$$<L^2> = \frac{\int_{-1}^{1} L^2 e^{-mf(L)/kT} dL}{\int_{-1}^{1} e^{-mf(L)/kT} dL} \approx \frac{kT/m}{f_2}.$$  Eq. (4)

Here the integrals have been evaluated in the Gaussian approximation, $f(L)=f_0 + f_2 L^2$, which is valid for disordered systems above the transition where $\langle L^2 \rangle$ is small. Inserting Eq. (4) into the internal energy per particle yields $\langle \varepsilon_m \rangle = -\tfrac{1}{2}\varepsilon_2 kT/f_2 m$, so that the average energy barrier for cooperative relaxation between energy minima is $m|\langle \varepsilon_m \rangle| = \tfrac{1}{2}\varepsilon_2 kT/f_2$. Note that the factor of $kT$ comes from normal thermal fluctuations about the free-energy minimum, which increase the barrier height with increasing temperature as the system fluctuates further from the energy maximum, see Fig. 1. This feature of a free-energy landscape is opposite to the usual picture of a potential-energy landscape, where thermal fluctuations about an energy minimum reduce the barrier height with increasing temperature. Using $E_N = m|\langle \varepsilon_m \rangle|$ as the activation energy in the Arrhenius law gives a characteristic relaxation time of $\tau_N \sim \exp[\tfrac{1}{2}\varepsilon_2/f_2]$. Indeed, the free-energy fluctuations eliminate both the explicit temperature- and size-dependence in $\tau_N$, so that the distribution of relaxation times is relatively independent of temperature, providing a common mechanism for approximate time-temperature superposition. However, the free-energy curvature depends implicitly on temperature via the difference between the internal energy and entropy terms, $f_2 = \tfrac{1}{2}(kT - u_2)$, so that $|\langle u_m \rangle| = u_2 kT/[m(kT - u_2)]$ as in Eq. (2), and the characteristic relaxation time obeys the VFT law $\tau_N \sim \exp[u_2/(kT - u_2)]$. Thus, both time-temperature superposition and the VFT law can be attributed to fluctuations in free energy of finite-sized regions.

The intrinsic heterogeneity can also be attributed to fluctuations in free energy. Again note the factor of $1/m$ in the internal energy per particle, $\langle \varepsilon_m \rangle = -\tfrac{1}{2}\varepsilon_2 kT/f_2 m$. This inverse size dependence comes from normal thermal fluctuations that decrease with increasing system size, yielding the classical mean-field result that the disordered phase has $<\varepsilon_m>=0$ as $m\to\infty$. Now consider a large, but finite system of say $m\sim10^{21}$ particles. If the system is homogeneous, so that all of these particles fluctuate coherently, then the average change in the interaction energy per particle due to fluctuations is effectively zero, $<\varepsilon_m>\sim-\varepsilon_2/10^{21}$; but if the system subdivides into $10^{20}$ independently fluctuating regions, each containing 10 particles, then the fluctuations lower the energy by a factor of $10^{20}$, $<\varepsilon_{10}>\sim-\varepsilon_2/10$. Although this energy reduction always favors separating a sample into the smallest possible regions, two mechanisms inhibit these regions from becoming the size of individual particles. Surface terms for small regions increase the subdivision energy, and cause interactions between the regions. However, the surface terms are negligible for sufficiently large regions, or for regions that fluctuate independently so that the surface terms are averaged,[24] as found by Monte Carlo simulations.[25] In any case, negligible coupling between fluctuating regions is consistent with measurements showing that the regions relax independently.

The more likely mechanism controlling region sizes is that subdividing a bulk sample also lowers its entropy. Thus, the true thermal equilibrium involves a balance between entropy and energy, which is found by minimizing the free energy. If there are no internal constraints on the size of the fluctuating regions, as expected inside bulk materials, then the appropriate free energy is the grand potential $\Omega(L)=m[f(L)-\mu]$, where $\mu$ is the chemical potential. $\Omega(L)$ is minimized by the generalized ensemble, with the partition function given by

$$\Gamma = \sum_{m=1}^{\infty} \int_{-1}^{1} e^{-m[f(L)-\mu]/kT} dL \qquad \text{Eq. (5)}$$

Note that the integral over $L$ corresponds to the usual sum over energies of a canonical ensemble partition function; but unlike the usual thermodynamic limit $m\to\infty$, we consider regions of all sizes $1 \leq m < \infty$. Moreover, this sum over all $m$ differs from the usual grand canonical ensemble because there is no volume dependence to fix the size of the regions. Thus, the sum results in a Legendre transform that converts the canonical ensemble into an ensemble of regions with all possible sizes. The generalized ensemble is not allowed in the usual thermodynamic limit, where at least one extensive parameter must be used to fix the size of the system. Indeed, the Gibbs-Duhem relation implies that the intensive variables cannot all be independent. However, the energy density of finite-sized regions is usually not constant; for example normal thermal fluctuations yield a $1/m$ term as in Eqs. (2) and (4). Furthermore, although the intensive parameters cannot all be independent if $m \to \infty$, the Gibbs-Duhem relation does not apply to finite systems, and the generalized ensemble is the only ensemble that minimizes the appropriate free energy for finite-sized regions inside a bulk sample without artificially fixing the size of the regions. The resulting distribution of sizes provides a common explanation for the distribution of dynamically correlated regions that are found in the response of viscous liquids.

## RESULTS AND DISCUSSIONS

Quantitative comparison of our Landau theory with experimental data requires a more careful consideration of the statistics of thermal fluctuations.[21,22] The basic idea is that each region of the sample can contain an unlimited number ($n$) of indistinguishable and independent fluctuations, so that the net partition function becomes $Y = \sum_{n=1}^{\infty} \Gamma^n / n!$. Figure 2 shows the temperature dependence of the characteristic response time (as determined from the peak dielectric loss frequency) of two glass-forming liquids.[39] On this type of plot an Arrhenius law with constant energies gives straight lines, whereas the data show curvature characteristic of the VFT law. A least-squares fit to the data yields $T_0 =$ 131 and 114 K with $B/T_0$=16.8 and 17.2 for glycerol and propylene glycol, respectively. The solid curves in Fig. 2 come from the temperature-dependent activation energy $E_N = |kT^2 \partial(\ln Y)/\partial T|$ using the Gaussian approximation to the free energy. Non-linear least squares fits to the data yield $\varepsilon_2/k =$ 121 and 104 K with $(f_0-\mu)/kT =$ 0.0337 and 0.0313 for glycerol and propylene glycol, respectively. Note that we use a constant value for $(f_0-\mu)/kT$, as is appropriate at $kT > u_2$ where the entropy term dominates the free energy, $f_0 \sim kT$. Also note that for these "strong" glass-forming liquids, using $E_N = |kT^2 \partial(\ln Y)/\partial T|$ instead of $E_N \propto \varepsilon_2 kT/(kT-\varepsilon_2)$ shifts the characteristic temperature by less than 10 %, so that the simple VFT law with $\varepsilon_2 = kT_0$ is a good approximation. In fact, adding the fourth-order term $f_4/kT = -1/12$ from an expansion of the binomial coefficient improves the agreement: $\varepsilon_2/k =$ 131 and 112 K with $(f_0-\mu)/kT =$ 0.0349 and 0.0323. Alternatively, the fourth-order term can be released as an adjustable parameter yielding $f_4/kT = -0.11 \pm 0.06$, which provides quantitative evidence that the entropy involves binary degrees of freedom, such as two level systems or competing local structures. Finally note that "fragile" glass-forming liquids, which have $T_0$ close to the glass transition temperature, show clear deviations from the VFT law that can be attributed to such higher-order terms in the entropy.[21,22]

The symbols in Fig. 3 show the measured dielectric loss of glycerol as a function of frequency at four different temperatures.[40] The solid curves come from fits to the data using our Landau theory for a distribution of relaxation times, Eq. (1), with $g_N = g_m g_n$ as the approximate probability of finding a region that contains $N=mn$ fluctuating particles. Here the distributions are $g_m = \int_{-1}^{1} e^{-m[f(L)-\mu]/kT} dL/\Gamma$ and $g_n = (\Gamma^n/n!)/Y$. The size-dependent part of the relaxation time, which comes from the internal-energy barrier at fixed temperature due to size-dependent fluctuations about the free-energy minimum, varies inversely proportional to size $\ln(\tau_N) \propto -1/N$, as described in Refs. 21 and 22.

Although it is always possible to project any physical system onto a single potential-energy landscape, here we argue that it is more instructive to expand the free-energy using a separate parameter for each independent region, at least for disordered systems with heterogeneity. The usual potential-energy landscape is assumed to be fixed, with all of the temperature dependence coming from kinetic energy added to the surface, so that empirical features such as the VFT law and time-temperature superposition are envisioned by sketching a sufficiently complex landscape. Here we have shown that normal thermal fluctuations about the free energy minimum yield energy barriers that are proportional to *kT*, effectively canceling the broadening inherent in the Arrhenius law to give approximate time-temperature superposition; and that the linear temperature dependence from the entropy term in the free energy provides a common explanation for VFT-like behavior. For disordered systems (broken curves in Fig. 1), the free-energy minimum occurs at *L*=0, where both energy and entropy are a maximum, so that the energy minima have negligible influence on the dynamics. In our theory, the free energy controls the amplitude of the thermal fluctuations, while the barrier that governs the time scale of the dynamics comes from the curvature about the internal energy maximum. Finally, in both the original Landau theory and standard potential-energy landscape pictures, the configuration of an entire sample is represented by a single point, which hides the heterogeneity that is crucial for understanding the dynamics.

Next we compare our Landau-like theory to the configurational entropy picture of Adam and Gibbs (AG).[41] We agree with AG that entropy plays a central role in the dynamics of disordered materials. We also agree that a bulk sample subdivides into an ensemble of small regions, and that these regions are weakly coupled to their environment so that they relax independently, as is now found by several measurement techniques. However, one difference is that AG assume that the dynamics is dominated by regions with a single critical size, $z^*$, so that there is a single relaxation time for the entire sample $\tau_{z^*} \propto \exp(\frac{z^* \Delta u}{kT})$; whereas we use the generalized ensemble to obtain a thermal equilibrium distribution of region sizes. Furthermore, the critical region sizes deduced from thermal properties using the AG theory are typically only about 4 particles near the glass transition,[42] significantly smaller than the region sizes that are measured directly.[28] Another difference is that AG assume that the activation energy per particle ($\Delta u$) is constant, independent of temperature; whereas we obtain the temperature dependences given in Eq. (1). AG obtain the VFT law from the critical size $z^* = \frac{NkT\ln(2)}{C(T-T_2)}$, which is based on the assumption that the configurational entropy dominates the dynamics, and that this entropy goes to zero linearly as the glass transition is approached. However, it has been shown that non-configurational degrees of freedom contribute significantly to the entropy of many materials;[43] and some investigators object to the idea that one component of a thermodynamic variable goes to zero linearly at a nonzero temperature, while other components are unaffected.[44] In Landau theory, the factor of $kT-\varepsilon_2$ in the denominator of Eq. (2) occurs naturally from the difference between the internal energy and entropy terms in the free energy, similar to the Curie-Weiss law from magnetism. Furthermore, minimizing the free energy using nanothermodynamics yields a size-dependent activation energy that provides a mechanism for the measured dynamical heterogeneity, and distribution of independently relaxing regions, which remain assumptions in the AG theory.

In summary, normal thermal fluctuations in the free energy of finite-sized regions yield an activation energy per particle, Eq. (2), that provides a unified physical picture for WLF-like behavior. The factor of $kT$ in the numerator cancels the explicit temperature dependence in the Arrhenius law to give approximate time-temperature superposition. The temperature dependence in the denominator, which comes from the entropy term in the free energy, provides a common mechanism for VFT-like behavior. The factor of $m$ in the denominator provides a general explanation for the distribution of independently relaxing regions in a bulk sample. Thus, a Landau-like theory with finite-size thermal effects provides a unified physical picture for the dynamics of disordered materials.

We thank R. Richert and P. Lunkenheimer for providing the original data used in Figs. 2 and 3. We also thank N. Bernhoeft, R. Richert, and G. H. Wolf for helpful comments. Partial support for this research was provided by NSF Grant No. DMR-0514592.

**FIG. 1.** Sketch of internal energy (solid curve) and grand potential per particle (broken curves) as a function of the dynamical order parameter, *L*. Quantitative values, from the fit to glycerol shown in Fig. 2, correspond to room temperature, the glass transition, and $T=\varepsilon_2/k$ from top to bottom, respectively. The internal-energy minima occur at $L=\pm 1$, but if $T\geq\varepsilon_2/k$ the free-energy minimum is at $L=0$. For a fixed free-energy curvature, normal thermal fluctuations about $L=0$ increase with increasing temperature, canceling the explicit temperature dependence in the Arrhenius law to give approximate time-temperature superposition. However, the thermal fluctuations and average potential-energy barrier tend to diverge as the curvature goes to zero at $T=\varepsilon_2/k$, yielding VFT-like behavior. The normal thermal fluctuations also increase with decreasing system size, so that subdividing a system into smaller regions lowers the net internal energy.

**FIG. 2.** Angell plot of the characteristic relaxation time as a function of inverse temperature. The symbols are from the measured peak dielectric loss of (○) glycerol and (□) propylene glycol from Ref. 39. The solid curves are from fits to the data using a fourth-order approximation of the free energy to obtain the temperature-dependent activation energy, $E_N=|kT^2\, \partial(\ln Y)/\partial T|$, as described in the text.

**FIG. 3.** Log-log plot of the frequency-dependent response of glycerol at four temperatures. The symbols are from measurements of dielectric loss from Ref. 40. The solid curves are from fits to the data using a distribution of sizes $g_N$ and size-dependent relaxation time $\ln(\tau_N) \sim -1/N$, as described in the text.

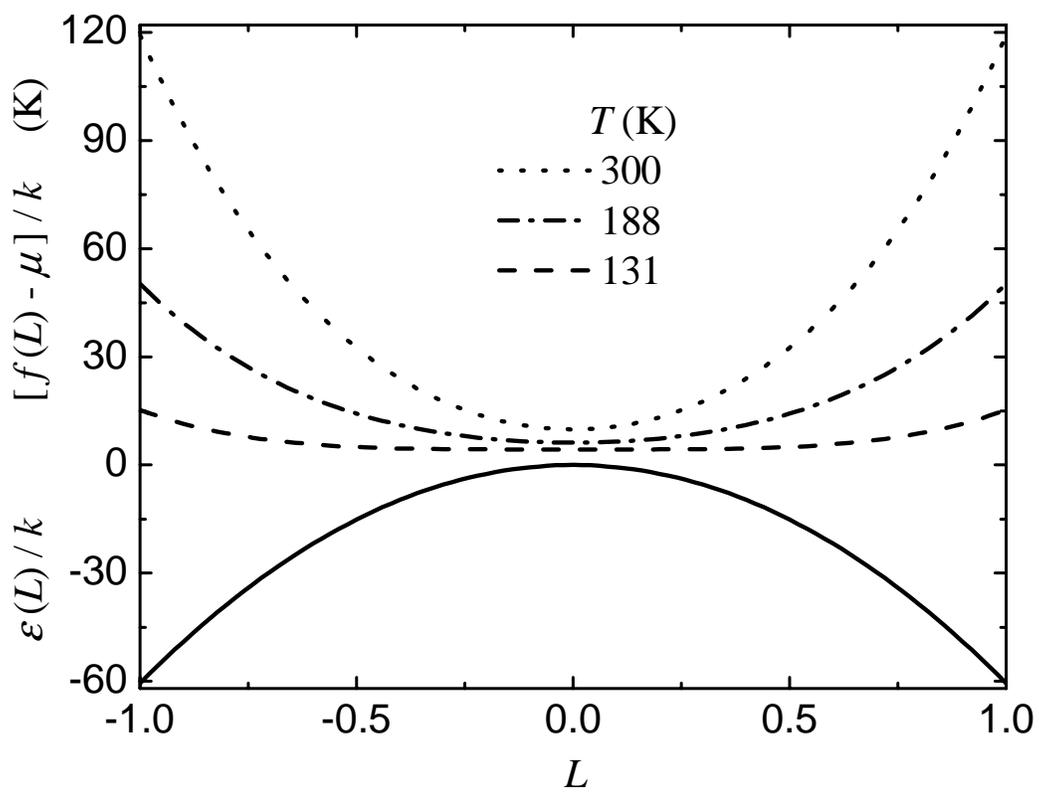

Figure 1

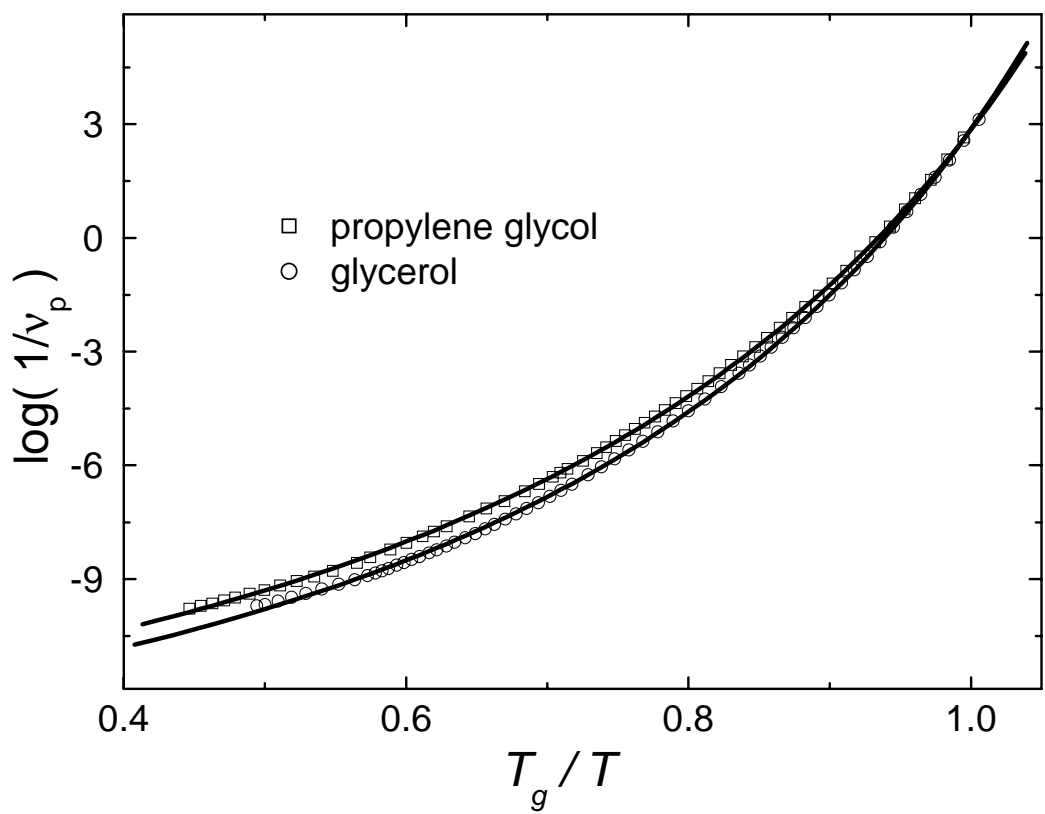

Figure 2

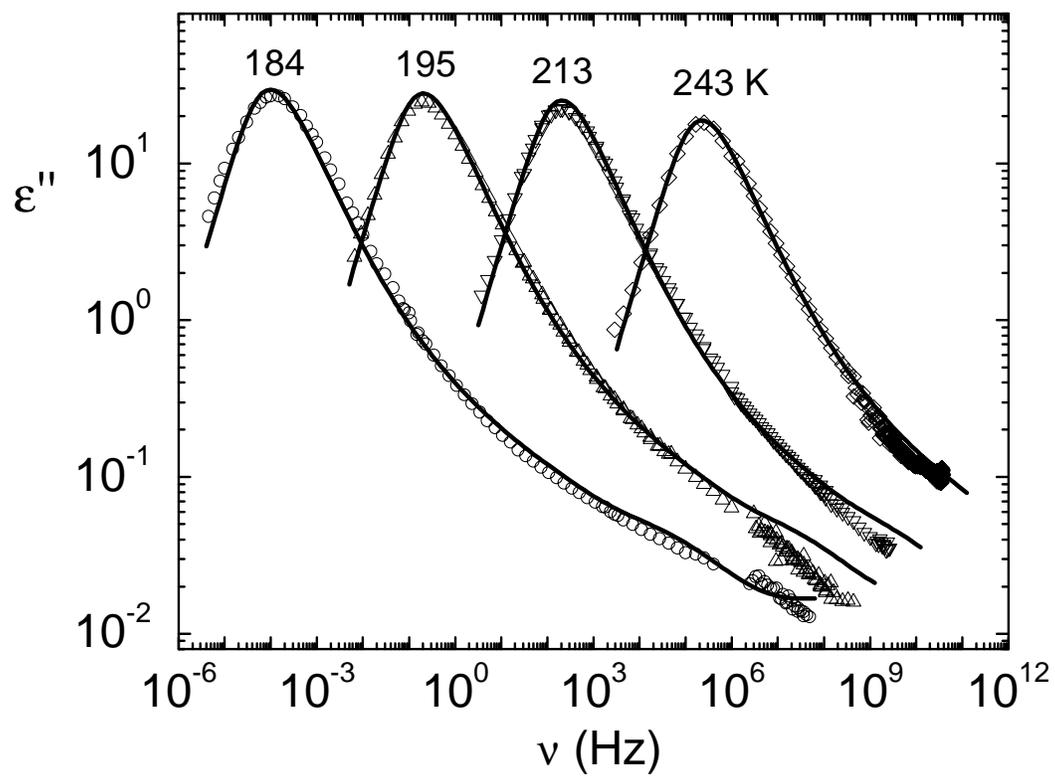

Figure 3